\documentclass{PoS}

\title{Scalar mesons and tetraquarks by means of lattice QCD}

\ShortTitle{Scalar mesons and tetraquarks by means of lattice QCD}

\author{European Twisted Mass Collaboration (ETMC):}

\author{\speaker{Marc Wagner} \\
        Goethe-Universit\"at Frankfurt am Main, Institut f\"ur Theoretische Physik, \\ $\quad$ Max-von-Laue-Stra{\ss}e 1, D-60438 Frankfurt am Main, Germany}

\author{Constantia Alexandrou, Mario Gravina \\
        Department of Physics, University of Cyprus, P.O.\ Box 20537, 1678 Nicosia, Cyprus \\
        Computation-based Science and Technology Research Center, Cyprus Institute, 20 Kavafi Street, \\ $\quad$ Nicosia 2121, Cyprus}

\author{Jan Oliver Daldrop, Carsten Urbach \\
        Helmholtz-Institut f{\"u}r Strahlen- und Kernphysik (Theorie) and Bethe Center for Theoretical \\ $\quad$ Physics, Universit{\"a}t Bonn, D-53115 Bonn, Germany}

\author{Mattia Dalla Brida \\
        School of Mathematics, Trinity College Dublin, Dublin 2, Ireland}

\author{Luigi Scorzato \\
        ECT$^\star$, Strada delle Tabarelle, 286, I-38123, Trento, Italy}

\author{Christian Wiese \\
        NIC, DESY Zeuthen, Platanenallee 6, D-15738 Zeuthen, Germany \\
        Humboldt-Universit\"at zu Berlin, Institut f\"ur Physik, Newtonstra{\ss}e 15, D-12489 Berlin, Germany}

\abstract{We study the light scalar mesons $a_0(980)$ and $\kappa$ using $N_f = 2+1+1$ flavor lattice QCD. In order to probe the internal structure of these scalar mesons, and in particular to identify, whether a sizeable tetraquark component is present, we use a large set of operators, including diquark-antidiquark, mesonic molecule and two-meson operators. The inclusion of disconnected diagrams, which are technically rather challenging, but which would allow us to extend our work to e.g.\ the $f_0(980)$ meson, is introduced and discussed.
}

\FullConference{Xth Quark Confinement and the Hadron Spectrum\\
                8-12 October 2012 \\
                TUM Campus Garching, Munich, Germany}

\begin{document}

% ********************
% ********************
% ********************
% ********************
% ********************

\section{Introduction}

The nonet of light scalar mesons formed by $\sigma \equiv f_0(500)$, $\kappa \equiv K_0^\ast(800)$, $a_0(980)$ and $f_0(980)$ is poorly understood. Compared to expectation all nine states are rather light and their ordering is inverted. For example in a standard quark antiquark picture the $a_0(980)$ states, which have $I = 1$, must necessarily be composed of two light quarks, e.g.\ $a_0(980) \equiv \bar{d} u$, while the $\kappa$ states with $I = 1/2$ are made from a strange and a light quark, e.g.\ $\kappa \equiv \bar{s} u$. Consequently, $\kappa$ should be heavier than $a_0(980)$, since $m_s > m_{u,d}$. In experiments, however, the opposite is observed, i.e.\ $m(\kappa) = 682 \pm 29 \, \textrm{MeV}$, while $m(a_0(980)) = 980 \pm 20 \, \textrm{MeV}$ \cite{PDG}. On the other hand in a four-quark or tetraquark picture the quark content could be $a_0(980) \equiv \bar{d} u \bar{s} s$ and $\kappa \equiv \bar{s} u (\bar{u} u + \bar{d} d)$ naturally explaining the observed ordering. Moreover, certain decay channels, e.g.\ $a_0(980) \rightarrow K + \bar{K}$, indicate that besides the two light quarks also an $s \bar{s}$ pair is present and, therefore, also support a tetraquark interpretation. A detailed discussion of light scalar mesons can be found e.g.\ in \cite{Jaffe:2004ph,Pelaez}.

In addition to the light scalar mesons there are also various tetraquark candidates among the heavy mesons, e.g.\ $D_{s0}^\ast(2317)$ and $D_{s1}(2460)$ (cf.\ e.g.\ \cite{Moir,Kalinowski:2012re}) or the charmonium states $X(3872)$, $Z(4430)$, $Z(4050)$, $Z(4250)$.

Here we report about the status of an ongoing long-term project with the aim to study possible tetraquark candidates from first principles using lattice QCD. Parts of this work have already been published \cite{Daldrop:2012sr,Alexandrou:2012rm}.

% ********************
% ********************
% ********************

\section{\label{SEC001}Lattice setup}

We use gauge link configurations with $N_f = 2+1+1$ dynamical quark flavors generated by the European Twisted Mass Collaboration (ETMC).

We have studied several ensembles with the same rather fine lattice spacing $a \approx 0.086 \, \textrm{fm}$. The ensembles differ in the volume $(L/a)^3 \times (T/a) = 20^3 \times 48 , \ldots , 32^3 \times 64$ and the unphysically heavy light quark mass corresponding to $m_\pi \approx 280 \, \textrm{MeV} \ldots 460 \, \textrm{MeV}$. Details regarding these gauge link configurations can be found in \cite{Baron:2008xa,Baron:2009zq,Baron:2010bv,Baron:2011sf,Baron:2010th,Baron:2010vp}.

Currently we ignore disconnected diagrams, which are technically rather challenging (cf.\ the outlook in section~\ref{SEC002}). An important physical consequence is that the quark number and the antiquark number are separately conserved for each flavor. Therefore, there is no mixing between $\bar{u} u$, $\bar{d} d$ and $\bar{s} s$ resulting in an $\eta_s$ meson with flavor structure $\bar{s} s$ instead of $\eta$ and $\eta'$.

Further lattice details and technicalities can be found in \cite{Alexandrou:2012rm}.

% ********************
% ********************
% ********************

\section{Four-quark creation operators}

In the following we focus on the $a_0(980)$ sector, which has quantum numbers $I(J^P) = 1(0^+)$. As usual in lattice QCD we extract the low lying spectrum in that sector by studying the asymptotic exponential behavior of Euclidean correlation functions
\begin{eqnarray}
C_{j k}(t) \ \ = \ \ \Big\langle (\mathcal{O}_j(t))^\dagger \mathcal{O}_k(0) \Big\rangle .
\end{eqnarray}
$\mathcal{O}_j$ and $\mathcal{O}_k$ denote suitable creation operators, i.e.\ operators generating the $a_0(980)$ quantum numbers, when applied to the vacuum state.

Assuming that the experimentally measured $a_0(980)$ with mass $980 \pm 20 \, \textrm{MeV}$ is a rather strongly bound four quark state, suitable creation operators to excite such a state are
\begin{eqnarray}
\label{EQN001} & & \hspace{-0.7cm} \mathcal{O}_{a_0(980)}^{K \bar{K} \textrm{\scriptsize{} molecule}} \ \ = \ \ \sum_\mathbf{x} \Big(\bar{s}(\mathbf{x}) \gamma_5 u(\mathbf{x})\Big) \Big(\bar{d}(\mathbf{x}) \gamma_5 s(\mathbf{x})\Big) \\
\label{EQN002} & & \hspace{-0.7cm} \mathcal{O}_{a_0(980)}^{\textrm{\scriptsize diquark}} \ \ = \ \ \sum_\mathbf{x} \Big(\epsilon^{a b c} \bar{s}^b(\mathbf{x}) C \gamma_5 \bar{d}^{c,T}(\mathbf{x})\Big) \Big(\epsilon^{a d e} u^{d,T}(\mathbf{x}) C \gamma_5 s^e(\mathbf{x})\Big) .
\end{eqnarray}
The first operator has the spin/color structure of a $K \bar{K}$ molecule ($\bar{s}(\mathbf{x}) \gamma_5 u(\mathbf{x})$ and $\bar{d}(\mathbf{x}) \gamma_5 s(\mathbf{x})$ correspond to a kaon $K$ and an antikaon $\bar{K}$ at the same position $\mathbf{x}$). The second resembles a bound diquark antidiquark pair, where spin coupling via $C \gamma_5$ corresponds to the lightest diquarks/antidiquarks (cf.\ e.g.\ \cite{Jaffe:2004ph,Alexandrou:2006cq,Wagner:2011fs}).

Further low lying states in this sector are the two particle states $K + \bar{K}$ and $\eta_s + \pi$. Suitable creation operators to resolve these states are
\begin{eqnarray}
\label{EQN003} & & \hspace{-0.7cm} \mathcal{O}_{a_0(980)}^{K + \bar{K} \textrm{\scriptsize{} two-particle}} \ \ = \ \ \bigg(\sum_\mathbf{x} \bar{s}(\mathbf{x}) \gamma_5 u(\mathbf{x})\bigg) \bigg(\sum_\mathbf{y} \bar{d}(\mathbf{y}) \gamma_5 s(\mathbf{y})\bigg) \\
\label{EQN004} & & \hspace{-0.7cm} \mathcal{O}_{a_0(980)}^{\eta_s + \pi \textrm{\scriptsize{} two-particle}} \ \ = \ \ \bigg(\sum_\mathbf{x} \bar{s}(\mathbf{x}) \gamma_5 s(\mathbf{x})\bigg) \bigg(\sum_\mathbf{y} \bar{d}(\mathbf{y}) \gamma_5 u(\mathbf{y})\bigg) .
\end{eqnarray}

% ********************
% ********************
% ********************

\section{Numerical results an their interpretation}

We first discuss numerical results for the ensemble with the smallest volume, $(L/a)^3 \times (T/a) = 20^3 \times 48$, which corresponds to a spatial extension of $L \approx 1.72 \, \textrm{fm}$. This ensemble is particularly suited to distinguish two-particle states with relative momentum from states with two particles at rest and from possibly existing $a_0(980)$ tetraquark states (two-particle states with relative momentum have a rather large energy because one quantum of momentum \\ $p_\textrm{\scriptsize min} = 2 \pi / L \approx 720 \, \textrm{MeV}$).

\begin{figure}[htb]
\input{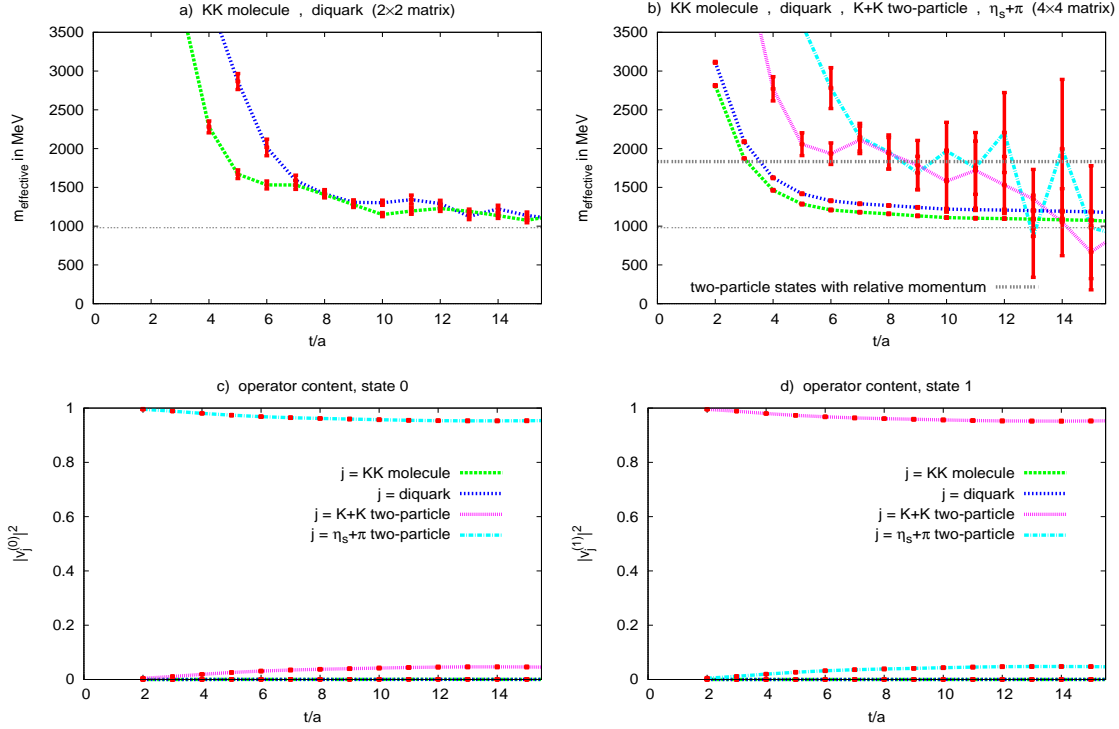}
\caption{\label{FIG001}$a_0(980)$ sector, $(L/a)^3 \times (T/a) = 20^3 \times 48$.
\textbf{a)}~Effective masses as functions of the temporal separation, $2 \times 2$ correlation matrix (operators: $K \bar{K}$ molecule, diquark-antidiquark, eqs.\ (3.2) and (3.3)).
\textbf{b)}~$4 \times 4$ correlation matrix (operators: $K \bar{K}$ molecule, diquark-antidiquark, two-particle $K + \bar{K}$, two-particle $\eta_s + \pi$, eqs.\ (3.2) to (3.5)).
\textbf{c)}, \textbf{d)}~Squared eigenvector components of the two low-lying states from \textbf{b)} as functions of the temporal separation.
}
\end{figure}

Figure~\ref{FIG001}a shows effective mass plots from a $2 \times 2$ correlation matrix with a $K \bar{K}$ molecule operator (\ref{EQN001}) and a diquark-antidiquark operator (\ref{EQN002}). The corresponding two plateaus are around $1100 \, \textrm{MeV}$ and, therefore, consistent both with the expectation for possibly existing $a_0(980)$ tetra\-quark states and with two-particle $K + \bar{K}$ and $\eta_s + \pi$ states, where both particles are at rest \\ ($m(K + \bar{K}) \approx 2 m(K) \approx 1198 \, \textrm{MeV}$; $m(\eta_s + \pi) \approx m(\eta_s) + m(\pi) \approx 1115 \, \textrm{MeV}$ in our lattice setup).

% ***
% ***
% ***

% ***
% ***
% ***

Increasing this correlation matrix to $4 \times 4$ by adding two-particle $K + \bar{K}$ and $\eta_s + \pi$ operators (eqs.\ (\ref{EQN003}) and (\ref{EQN004})) yields the effective mass results shown in Figure~\ref{FIG001}b. Two additional states are observed, whose plateaus are around $1500 \, \textrm{MeV} \ldots 2000 \, \textrm{MeV}$. From this $4 \times 4$ analysis we conclude the following:
\begin{itemize}
\item We do not observe a third low-lying state around $1100 \, \textrm{MeV}$, even though we provide operators, which are of tetraquark type as well as of two-particle type. This suggests that the two low-lying states are the expected two-particle $K + \bar{K}$ and $\eta_s + \pi$ states, while an additional stable $a_0(980)$ tetraquark state does not exist.

\item The effective masses of the two low-lying states are of much better quality in Figure~\ref{FIG001}b than in Figure~\ref{FIG001}a. We attribute this to the two-particle $K + \bar{K}$ and $\eta_s + \pi$ operators, which presumably create larger overlap to those states than the tetraquark operators. This in turn confirms the interpretation of the two observed low-lying states as two-particle states.

\item To investigate the overlap in a more quantitative way, we show the squared eigenvector components of the two low-lying states in Figure~\ref{FIG001}c and Figure~\ref{FIG001}d (cf.\ \cite{Baron:2010vp} for a more detailed discussion of such eigenvector components). Clearly, the lowest state is of $\eta_s + \pi$ type, whereas the second lowest state is of $K + \bar{K}$ type. On the other hand, the two tetraquark operators are essentially irrelevant for resolving those states, i.e.\ they do not seem to contribute any important structure, which is not already present in the two-particle operators. These eigenvector plots give additional strong support of the above interpretation of the two observed low lying states as two-particle states.

\item The energy of two-particle excitations with one relative quantum of momentum can be estimated according to
\begin{eqnarray}
m(1 + 2,p = p_\textrm{\scriptsize min}) \ \ \approx \ \ \sqrt{m(1)^2 + p_\textrm{\scriptsize min}^2} + \sqrt{m(2)^2 + p_\textrm{\scriptsize min}^2} \quad , \quad p_\textrm{\scriptsize min} \ \ = \ \ \frac{2 \pi}{L} .
\end{eqnarray}
Inserting the meson masses corresponding to our lattice setup, $m(K) \approx 599 \, \textrm{MeV}$, \\ $m(\eta_s) \approx 774 \, \textrm{MeV}$ and $m(\pi) \approx 341 \, \textrm{MeV}$, yields $m(K + \bar{K},p = p_\textrm{\scriptsize min}) \approx 1873 \, \textrm{MeV}$ and \\ $m(\eta_s + \pi,p = p_\textrm{\scriptsize min}) \approx 1853 \, \textrm{MeV}$. These numbers are consistent with the effective mass plateaus of the second and third excitation in Figure~\ref{FIG001}b. Consequently, we also interpret them as two-particle states.
\end{itemize}

We obtained qualitatively identical results, when varying the light quark mass and the spacetime volume, as discussed in section~\ref{SEC001}. Corresponding plots are shown in \cite{Alexandrou:2012rm}.

Using exactly the same techniques, i.e.\ four-quark operators of tetraquark and of two-particle type, we also studied the $\kappa$-sector (for details cf.\ \cite{Alexandrou:2012rm}). Again we find no sign of any four-quark bound state besides the expected two-particle spectrum (in this case $K + \pi$ states). Note that this result is in contradiction to a very similar recent lattice study of the $\kappa$ meson \cite{Prelovsek:2010kg}, where an additional low lying four-quark bound state has been observed.

% ********************
% ********************
% ********************

\section{\label{SEC002}Inclusion of singly disconnected diagrams}

As mentioned in section~\ref{SEC001} disconnected diagrams have been ignored for the results presented so far. In this section we briefly discuss our strategy for computing such diagrams for specific four-quark operators.

\begin{figure}[htb]
\input{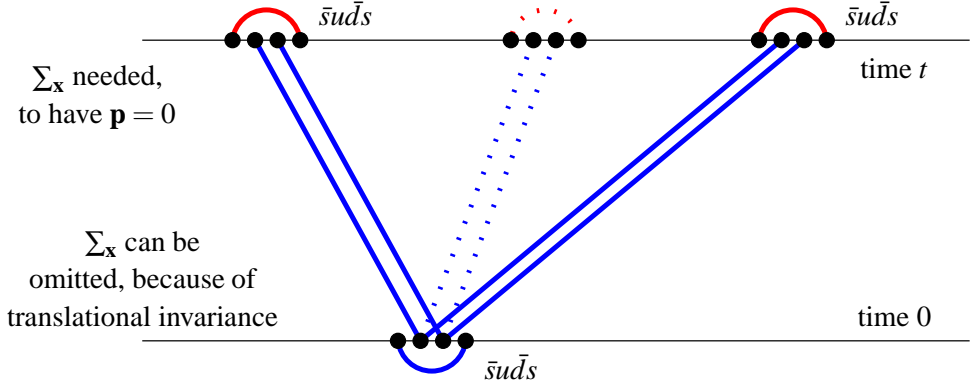}
\caption{\label{FIG002}The singly disconnected diagram of the $a_0(980) \equiv \bar{s} s \bar{d} u$ correlator.}
\end{figure}

For correlation functions of four-quark operators with flavor structure $\bar{q}_1 q_1 \bar{q_2} q_3$ a so-called singly-disconnected diagram has to be computed. Tetraquark candidates with this flavor structure are e.g.\ the previously discussed $a_0(980) \equiv \bar{s} s \bar{d} u$ (cf.\ Figure~\ref{FIG002}) or \\ $D_{s0}^\ast(2317) , D_{s1}(2460) \equiv (\bar{u} u + \bar{d} d) \bar{c} s$.

While for connected four-quark diagrams (i.e.\ all four quark propagators connect the timeslices at time $0$ and time $t$) standard point-to-all propagators can be used, applying exclusively such propagators is not possible in practice for singly disconnected diagrams. The reason is that one has to include a sum over space, $\sum_\mathbf{x}$, at least on one of the two timeslices (wlog.\ at time $t$ in Figure~\ref{FIG002}), to project to zero momentum. This in turn requires a all-to-all propagator of quark flavor $q_1$ on that timeslice, due to $\sum_\mathbf{x} \bar{q}_1(\mathbf{x},t) q_1(\mathbf{x},t) \ldots$ (the $s$ quark propagator at timeslice $t$ represented by the solid red line in Figure~\ref{FIG002}).

Since all-to-all propagators are prohibitively expensive to compute, they are typically estimated stochastically. While using a single stochastic propagator for a specific diagram typically results in a favorable or at least acceptable signal-to-noise ratio, using a larger number of such propagators drastically increases statistical errors. Therefore, we decided for the following strategy: three quark propagators (the $q_1$-loop at timeslice $0$ and the $q_2$ and $q_3$ propagators connecting the timeslices $0$ and $t$) are realized by exact point-to-all propagators, while the remaining propagator (the $q_1$-loop at timeslice $t$) is estimated stochastically, using random $Z_2 \times Z_2$ timeslice sources.

First numerical tests of this strategy performed for both $a_0(980)$ and $D_{s0}^\ast(2317)$ have been promising in the sense that the statistical errors of the singly disconnected diagrams for the molecule type operator (\ref{EQN001}) are of the same order of magnitude as the statistical errors of the corresponding connected diagrams, when investing a comparable amount of HPC resources.

% ***
% ***
% ***

% ***
% ***
% ***

% ********************
% ********************
% ********************

\section{Conclusions and future plans}

We have studied the $a_0(980)$ and the $\kappa$ channel by means of $N_f = 2+1+1$ flavor lattice QCD using four-quark operators of molecule, diquark and two-particle type. Besides the expected two-particle spectrum (two essentially non-interacting pseudoscalar mesons) no indication of any additional low lying state, in particular no sign of a four-quark bound state could be observed. This suggests that both the $a_0(980)$ and $\kappa$ meson have either no sizeable tetraquark component or they are rather weakly bound unstable states.

To investigate the latter one needs to study the volume dependence of the two-particle spectrum in the corresponding sectors (``L\"uscher's method'', cf.\ e.g.\ \cite{Luscher:1986pf,Luscher:1990ux,Luscher:1991cf}). Such computations are very challenging using lattice QCD, but first results have recently been published (cf.\ \cite{Lang:2012sv,Mohler:2012na}). We plan to perform similar computations with our setup in the near future.

Moreover, certain possibly present systematic errors need to be studied, quantified and removed: (1) disconnected diagrams have to be computed and included (cf.\ section~\ref{SEC002}); (2) lattice discretization errors and the continuum limit has to be studied; (3) computations at even lighter and, therefore, more realistic $u/d$ quark masses would be desirable.

% ********************
% ********************
% ********************

\section*{Acknowledgments}

M.W.\ acknowledges support by the Emmy Noether Programme of the DFG (German Research Foundation), grant WA 3000/1-1. M.G. acknowledges support by the Marie-Curie European training network ITN STRONGnet grant PITN-GA-2009-238353. M.D.B. is currently funded by the Irish Research Council, acknowledges support by STRONGnet and the AuroraScience project, and is grateful for the hospitality at ECT* and the University of Cyprus, where part of this work was carried out. L.S. acknowledges support from the AuroraScience project funded by the Province of Trento and INFN.

This work was supported in part by the Helmholtz International Center for FAIR within the framework of the LOEWE program launched by the State of Hesse and by the DFG and the NSFC through funds provided to the Sino-German CRC 110 ``Symmetries and the Emergence of Structure in QCD''.

Part of the computations presented here were performed on the Aurora system in Trento.

% ********************
% ********************
% ********************

% ********************
% ********************
% ********************

\end{document}